# Imaging of the ($Mn^{2+}3d^{5}$ + hole) complex in GaAs by Cross-sectional Scanning Tunneling Microscopy

A.M. Yakunin[1*], A.Yu. Silov[1], P.M. Koenraad[1], W. Van Roy[2], J. De Boeck[2], J.H. Wolter[1]
[1]COBRA Inter-University, Eindhoven University of Technology, P.O.Box 513, NL-5600MB Eindhoven, The Netherlands
[2]IMEC, Kapeldreef 75, B-3001 Leuven, Belgium
*Corresponding author: Phone: +31(040)2474190 Fax: +31(040)2461339 Email: a.m.yakunin@tue.nl

**Abstract**: We present results on the direct spatial mapping of the wave-function of a hole bound to a Mn acceptor in GaAs. To investigate individual Mn dopants at the atomic scale in both ionized and neutral configurations, we used a room temperature cross-sectional scanning tunneling microscope (X-STM). We found that in the neutral configuration manganese manifests itself as an anisotropic cross-like feature. We attribute this feature to a hole weakly bound to the Mn ion forming the ($Mn^{2+}3d^{5}$ + *hole*) complex.

## 1. Introduction

By now it is generally accepted that ferromagnetism in diluted magnetic semiconductors such as $Ga_{1-x}Mn_xAs$ is driven by the valence band states. A detailed investigation of the hole distribution around magnetic dopants is therefore of essential importance [1].

Deep acceptors in III/V semiconductors such as $Mn_{Ga}$ have been studied intensively for at least 30 years with a great variety of different techniques such as piezo-spectroscopy, hot photoluminescence, EMR etc, X-STM. However, hardly any information has been obtained on the electronic configuration at the atomic scale. Scanning tunneling microscopy is an ideal technique for the spatial investigation of complicated electronic structures, such as long-range interaction between an impurity and the host crystal. STM study of shallow acceptors such as Zn and Cd reveals that the behavior of doping atoms is still not fully understood [3-4]. Recently, manganese acceptors in their ionized state in GaAs have been studied by X-STM [2]. In this work we present a detailed investigation at atomic scale where we have used the STM tip as a tool to manipulate the manganese acceptor charge state $A^-/A^0$. We have imaged the Mn dopant by X-STM in both these charge states.

## 2. Samples

The measurements were performed on MBE grown samples. GaAs layers of 1200 nm doped with Mn at a concentration of about $3*10^{18}$ $cm^{-3}$ were grown on an intrinsic (001) GaAs substrate were. To prevent appearance of the structural defects such as arsenic antisites the samples were grown at a temperature of about 580 °C on. The Hall concentration of holes at room temperature was about $1.5*10^{17}$ $cm^{-3}$. The density of holes at room temperature is smaller than the doping concentration due to the limited activation of the carriers (activation energy of the Mn acceptor in GaAs is about 113 meV). The samples were non-conducting below 77 K. The X-STM experiments were performed at room temperature on an in-situ cleavage induced mono-atomically flat (110) surface in UHV ($P < 2*10^{-11}$ torr).

## 3. Experiment

In our experiment the low concentration of manganese is essential for the following two reasons: 1) it allows one to study the influence of a single Mn dopant on the host crystal electronic structure without an interaction with neighboring Mn dopants; 2) at such low concentration the tip-induced band bending extends as far as 10-15 nm into the semiconductor. The effect of the local perpendicular field from the STM tip is therefore small as compared to the binding energy of a Mn acceptor. At the same time the lateral extension of the band bending is much larger then Bohr radius. The tip-induced band bending can be manipulated by varying the sample-tip bias. In the case of a negative sample-bias (Fig. 1a), the Mn acceptor state is pulled below the

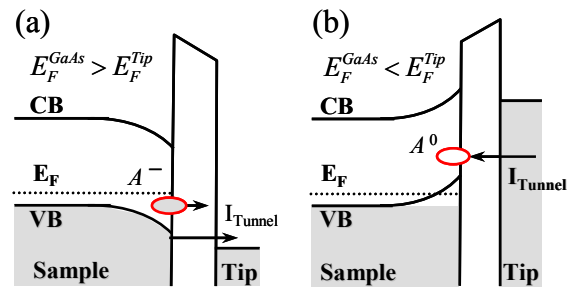

Fig. 1: Energy band diagram illustrating the tunneling process between tungsten tip and GaAs (110) surface in the presence of a tip induced band bending: (a) negative sample bias, filled states tunneling; (b) positive sample bias, empty states tunneling.

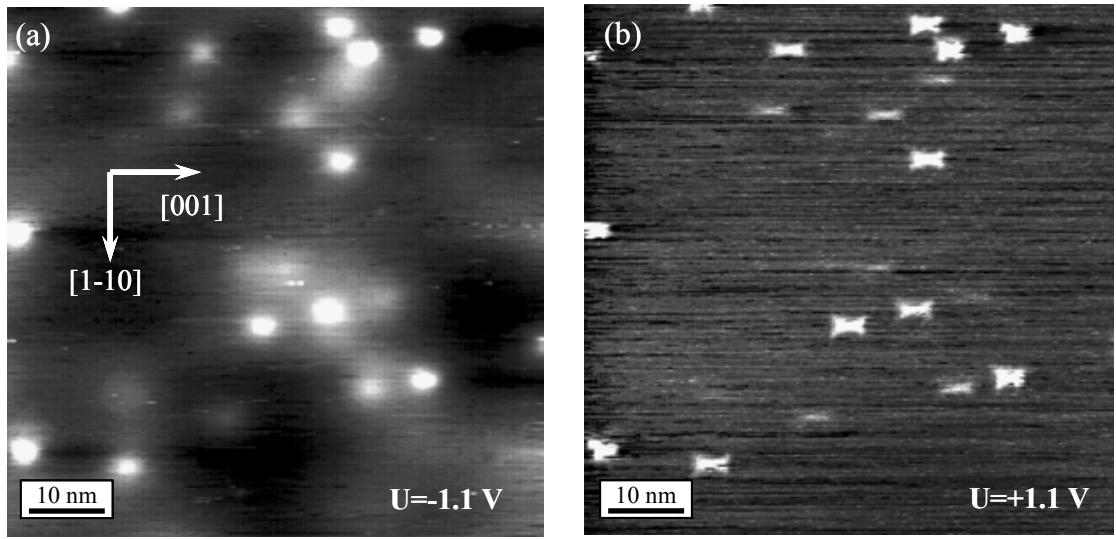

Fig. 2: Constant-current STM images 70x70 nm² of the same area. The Mn dopants are either in their (a) ionized (sample voltage is -1.1 V) or (b) neutral (sample voltage is +1.1 V) charge state. Both images display electronic contrast. In the image (a) the contrast is dominated by Coulomb field influence of the negatively charged $Mn_{Ga}$ ions on the neighboring states available for the tunneling. (b) The conductance is very poor since there is small amount of states available for the tunneling at this set point. The bright anisotropic feature appears as soon as the acceptor state is available for the tunneling. The brightest ones have huge electronic contrast, as big as 9 Å. The details of the feature are still visible without atomic resolution.

Fermi-level of the sample, thus Mn becomes ionized. This is similar to the case of energetically shallow acceptors like Be or Zn previously studied with X-STM [3-5]. At a negative bias such impurities appear as isotropic round elevations, which is a direct consequence of the influence of the negative ion and corresponding Coulomb potential on the number of states available for tunneling. At positive bias (Fig. 1b) acceptors are above the Fermi level and are in their neutral state. Under these conditions the hole occupying acceptor state is available for tunneling and it might be possible to map its density of states.

The electronic structure around Mn dopant was investigated for a wide range of applied biases by operating the STM in the constant-current mode. Three main regions in the electronic spectrum can be selected.

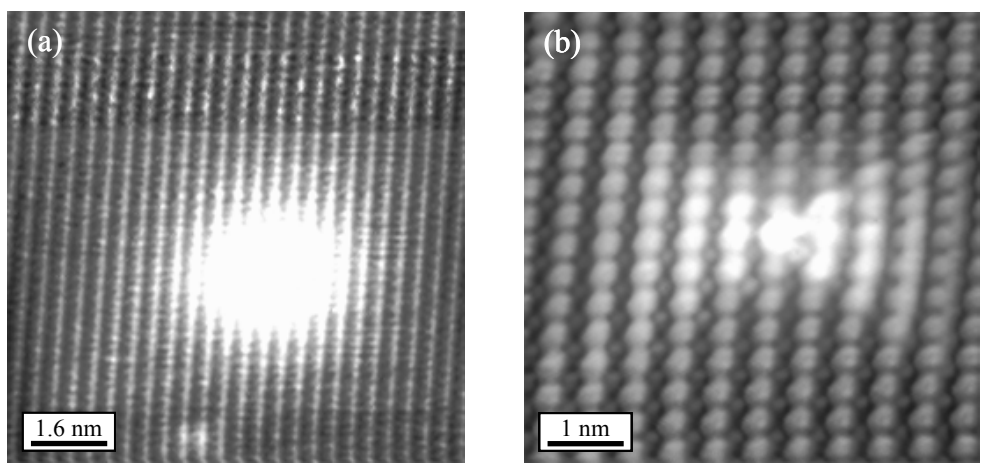

Fig. 3: (a) 10x10 nm² constant-current image taken at -0.7 V shows round isotropic elevation induced by ionized manganese. (b) 6x6 nm² constant-current image taken at +0.6 V. Due to the particular tunneling conditions a distinctive resolution is achieved. Note, that in the right part of the cross-like feature some apparent displacement of atomic rows occurs.

1) Negative sample-bias (filled states imaging) U < 0 V, Fig. 2a: as expected, the Mn impurity shows up in the STM image as a round isotropic elevation with a size and electronic contrast similar to what has been previously observed with STM on the ionized shallow acceptors in GaAs (Be, Cd, Zn). 2) Positive sample-bias (empty states imaging) 0.6 V < U < 1.5 V, Fig. 2b: In the region of the GaAs band-gap above the flat band-potential ($U_{FB}$=0.6 eV) the electronic contrast is dominated by a cross-like elevation. 3) Positive sample-bias U > 1.5 V. At the edge of the conduction band $U_C$=1.5 V, electronic contrast of the cross-like feature induced by Mn dopant is very weak and completely disappears at higher voltages when conduction band empty states dominate the tunnel current (see Fig. 5).

## 4. Discussion

Depending on the concentration and host crystal, substitutional $Mn_{III}$ can be found in three different electronic configurations. Firstly, it can be an ionized acceptor $A^-$ in the $Mn^{2+}3d^5$ electronic state. Secondly, in the neutral acceptor state $A^0$ formed by a negatively charged core weakly binding a valence hole forming a ($Mn^{2+}3d^5$ + *hole*) complex. Thirdly, a neutral configuration $3d^4$ can occur if a hole enters into the *d*-shell of Mn [7]. Mn interstitial can also act as a double donor. The electronic configuration of Mn dopants strongly influences the magnetic properties of the crystal.

The concentration of the dopants we observe in the STM corresponds to the intentional $3*10^{18}$ cm$^{-3}$ doping level. All of the observed dopants can be found either in the ionized $A^-$ or neutral $A^0$ charge state depending on the applied sample bias, as seen from the figures 2a and 2b.

We suggest that the localized cross-like features are the images of the holes confined by Mn acceptors. Each such a hole resides in a relatively deep state situated in the gap of GaAs at 100±30 meV above the valence band edge as seen from the shift of the branch of the I(V) at low negative voltages (see Fig. 5). This is in agreement with the ionization energy of Mn acceptor $E_a$=113 meV. At higher positive voltages the feature has weaker electronic contrast and disappears when the conduction band empty states dominate the tunnel current. This is the direct evidence that no extra charge is confined by the Mn studied under these conditions. The dopant thus is a neutral acceptor $A^0$. Because the confined hole extends over more than 5 bi-layers we conclude that it is loosely bound to the Mn ion forming the ($Mn^{2+}3d^5$ + *hole*) complex. This electronic configuration is in agreement with photoluminescence under uniaxial stress and Raman-scattering experiments [7-8]. The electronic contrast of the cross-like feature can be increased when tunneling at lower positive voltages in the region of the band gap. Such a tunneling appears to be possible in the case of p-type semiconductor due to upward band bending in the presence of the tip.

Now we turn to a discussion about the geometry of the observed feature. The cross-like feature extends 5-6 bi-layers around Mn dopant, which is about 6 nm in total along the [001] direction. We observe features in 5 different sub-surface layers. This distinction is mainly based on contrast difference, where Mn atoms situated closer to the surface are brighter. The size does not vary much with the depth.

The orientation of these cross-like features is the same for all observed manganese atoms and is geometrically linked to the host crystal. At any depth the mirror plane of the feature is (1-10). The image is weakly asymmetric with respect to [1-10] surface direction. We believe that the symmetry of the (110) surface is a cause of this distortion (see Figs. 3b and 4). The distortion becomes more pronounced when the imaged dopant is situated closer to the surface. In the figure 4 the left-hand part extends further away than the right-hand one. The orientation of the larger part is the same as of the triangular features induced by Zn and Cd dopants in GaAs. In the area of the smaller part there is a considerable atomic corrugation change (see Fig. 3b). This apparent shift of atomic rows in the [001] direction can be as much as 2.5 Å even when the dopant is situated as deep as in 3-4 sub-surface layers. Since this corrugation change appears only at low positive voltages and is not observed at higher voltages we conclude that it has pure electronic origin and is not related to a reconstruction or considerable lateral displacement of atoms.

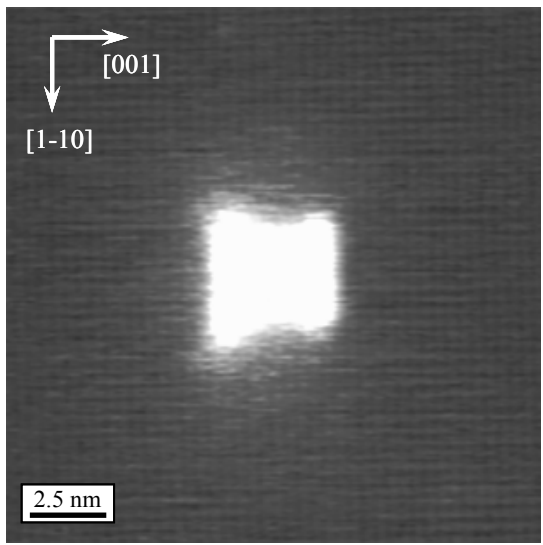

Fig. 4. 18x18 nm$^2$ constant-current image taken at +0.9 V shows spatial anisotropy of the feature induced by manganese in the neutral configuration.

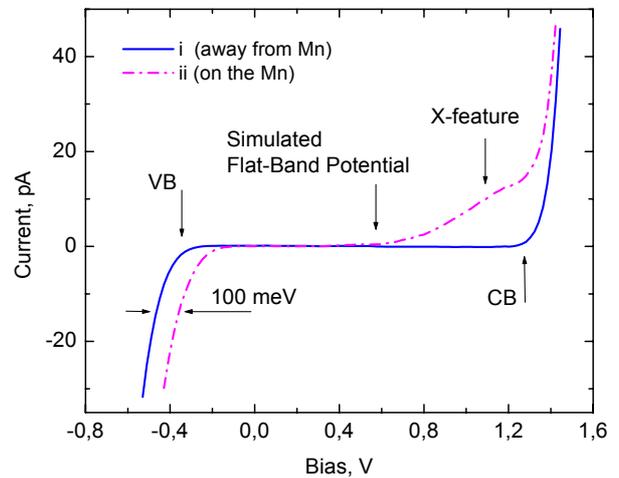

Fig. 5. Local Tunneling Spectroscopy [6]. The Set Point has been selected at +1.6 V and 100 pA. At each point of the image an I(V) spectrum is taken. At this bias the electronic contrast of the cross-like feature is almost absent, thus the tip-sample distance remains constant for each spectra. Spectrum (i), taken at a pure GaAs (110) surface away from Mn, displays a band gap of about 1.5 eV; (ii) taken on the Mn. In the area of the cross-like feature an extra current-channel appears in the band gap at low positive voltages. Imaging at this set point yields high contrast images.

## 5. Conclusion

We studied the electronic structure of a single Mn acceptor in GaAs in its both the ionized and neutral configurations at atomic scale. The neutral configuration of a single $Mn_{Ga}$ in GaAs was identified as a complex of a negative core and a weakly bound valence-hole ($Mn^{2+}3d^{5}$ + *hole*). We have mapped the density of states of this hole in the vicinity of the cleavage-induced surface. Our experiments reveal an anisotropic spatial structure of the hole, which appears in STM image as a cross-like feature.

**Acknowledgements**

This work was supported by the Dutch Foundation for Fundamental Research on Matter (FOM), Belgian Fund for Scientific Research Flanders (FWO) and EC GROWTH project FENIKS (GR5D-CT-2001-00535).